\documentclass[12pt]{article}

\usepackage{ucs}
\usepackage[utf8]{inputenc}
\usepackage{amsmath}
\usepackage{amsfonts}
\usepackage{amssymb}
\usepackage{amsthm}
\usepackage{rotating}
\usepackage{xcolor}
\usepackage[english]{babel}
\usepackage[T1]{fontenc}
\usepackage{graphicx}
\usepackage{float}
\usepackage{hyperref}
\usepackage{cuted}
\usepackage{verbatim}
\usepackage{subfigure}
\usepackage{multirow}
\usepackage{bbold}
\usepackage{slashed}
\usepackage{indentfirst}
\usepackage{tcolorbox}
\usepackage{xcolor}
\usepackage{placeins}
\usepackage[a4paper, total={7in, 10in}]{geometry}

\usepackage{multicol}
\usepackage{fmtcount}
\usepackage{diagbox}
\usepackage{caption}
\usepackage{subcaption}
\newcommand{\bpsi}{\bar{\psi}}

\begin{document}

\title{ \bf Symmetry breaking effects in 
pion  couplings to constituent quark  currents
}

 \author{ Fabio L. Braghin 
\\
{\normalsize Instituto de F\'\i sica, Federal University of Goias}
\\
{\normalsize Av. Esperan\c ca, s/n,
 74690-900, Goi\^ania, GO, Brazil}
}

\maketitle

\begin{abstract} 
Pseudoscalar and axial neutral and charged pion-constituent quark coupling constants
are investigated with nondegenerate quark masses in different
kinematical points, off shell and on shell pions and constituent quarks. 
By considering a large quark mass expansion of a quark 
determinant in the presence of
 local pion field and  of constituent quark background currents,
 gluonic effects are considered by means
of an effective gluon propagator that dresses quark currents.
For the neutral pion, mixing effects are introduced by 
means of the pion mixing to states $P_0$ and $P_8$,
that give rise to the $\pi^0-\eta-\eta'$ meson mixing,
and mixing of quark currents via corresponding 
mixing interactions.
The relative behavior of charged and neutral pion coupling constants
to quarks may be nearly the same - in the framework of 
the constituent quark model - as the pion-nucleon coupling constants
if mixings are introduced.
A very  small pion coupling to strange quark current is also obtained.
The dependence of the positive and negative pion-constituent quark
 coupling constant on the non-degeneracy of quark masses, for emission and absorption processes, 
 is identified.
 \end{abstract}

\section{Introduction}
\label{intro}

Pions, as the (quasi)Goldsone bosons of Strong Interactions,
have small masses as compared to baryons.
As a consequence, pion couplings to other hadrons, especially to
nucleons, have importance for a large number of low 
and intermediary energy processes.
In spite of their small masses,  they are well described in the framework of 
the constituent quark model (CQM) 
\cite{CQM,CQM2} 
in which a large binding energy
is required.
In part of these models, 
dynamical chiral symmetry breaking (DChSB)  plays a role of 
one of the mechanisms of mass generation
and contributes for  large quark (effective) masses, 
although other mechanisms may contribute
\cite{massgeneration1}.
In most versions of the CQM, mesons interact with constituent quarks belonging to
a baryon. 
These couplings must lead to the 
measurable meson-baryon coupling constants
that parameterize such interactions phenomenologically.
Several of the corresponding coupling constants 
  might follow  well known relations usually based in 
chiral symmetry and its dynamical breakdown
such as 
Goldberger Treiman relation (GTR)  
and partially conserved axial current
 \cite{chengLi-IZ,PCAC-gunnar}.
With the improvement of theoretical models and approaches
and the increase of experimental precision,
 it became important to understand further 
effects of isospin and flavor symmetry breakings (ISB and FSB).

There are several possible pion couplings to nucleons 
(or 
constituent quarks)
with differerent Lorentz structures  among which
the axial  and the pseudoscalar couplings are usually the most relevant ones.
The corresponding coupling constants,
 and  form factors,
parameterize 
 processes such as pion emission or absorption, besides
being responsible for the Yukawa meson exchange
among other processes.
Isospin symmetry breaking effects in the pseudoscalar 
pion coupling to the nucleon have been investigated recently 
within different approaches \cite{granada,bochum}
 leading to  similar conclusions 
 that these ISB effects must be very  small.
According to these works,
 the pseudoscalar neutral pion-neutron  coupling constant 
is larger than 
 the pseudoscalar neutral pion-proton  coupling constant.
However, it may be either 
slightly
 larger or considerably smaller  than the charged  pion coupling constant.
Nevertheless, the resulting overall small ISB
 is consistent with the fact that isospin symmetry breaking 
in flavor SU(2) QCD is  ruled by   up and down quark 
mass difference that is very small \cite{leutwyler,Donoghue,PDG}.
Other related consequences have been exploited in the literature, for example in \cite{granada,ISB-piN}.

In several versions of 
the CQM, the pion axial coupling 
is usually considered to be $g_A=1$, 
including in large Nc EFT proposed by 
Weinberg \cite{weinberg-2010,EPJA-2016}, being the axial charge 
directly related to 
the nucleon spin \cite{RMP2013}.
In the CQM, meson couplings to baryons, for example, are decomposed in
meson couplings to constituent quarks
 \cite{LeYaouanc1973}.
Recently, a
 quark determinant was analyzed in the presence 
of local meson fields and constituent (background) quark currents \cite{EPJA-2016,PRD-2019,PLB-2016}.
In a large quark mass expansion,  
both pseudoscalar and axial couplings appear, besides 
other leading scalar and vector couplings with the possibility of higher 
order (non-leading) 
 effective  couplings.
In this approach,
by starting with a leading term of the QCD effective quark action
for quark-antiquark interaction, by means of a (nonperturbative)
gluon exchange, quark currents are split into quantum quark currents and 
background quark currents, that are dressed by 
components of a  non perturbative gluon propagator  leading to 
constituent quark currents. 
Quantum quark currents
 lead to meson fields,  and to  the chiral condensate,
  by means of the
auxiliary field method.
Mesons structure and dynamics 
arise in flavor multiplets of quark-antiquark components
similarly to previous developments
\cite{ERV,GCM,meissner,chinese}.
In this approach,
pseudoscalar meson mixings  due to FSB
have been identified 
\cite{feldmann-review}
being possible to add 
to other known mechanism for meson mixing by means of
instanton-induced interactions
 \cite{instantons-ind}.
 In the present work, flavor mixing are investigated 
in two ways, by means of mixings for $\pi^0-\eta-\eta'$ mesons
and flavor mixing at the level of quark flavor currents
in effective quark-antiquark interactions, 
$G_{ij} (\bar{\psi} \Gamma \lambda_i \psi ) (\bar{\psi} 
\Gamma \lambda_j \psi )$ for $\Gamma = i\gamma_5$,
that have been
derived by a similar method
\cite{PRD-2021,JPG-2022}.

In this work, the axial and pseudoscalar pion-constituent quark 
  coupling 
constants
  will be derived from a quark determinant and investigated 
by considering nondegenerate up and down    quark masses.
The longstanding problem of establishing relations between
quark and hadron level quantities
are usually firstly envisaged by means 
of group theoretical framework
although the whole picture might become more intrincated
  due to several involved issues such as 
confinement and to the dynamical character of 
Bethe-Salpeter and Faddeev type equations for bound quarks.
Although this whole project may depend on the confinment 
mechanism and related properties, some aspects of
  hadron   are seemingly independent of it
according to phenomenology in which the CQM finds  
support for many hadrons.
Afterall, color is confined but
flavor is not.
With this program, the
 validity of the CQM description should be better
defined and refined by investigating its consequences such as 
fine tuning (ISB or FSB) effects in meson couplings.

Therefore, in the present work, an investigation of the ISB (and FSB) in  
constituent quark-pion interactions
 is proposed by means of 
the pseudoscalar and axial pion couplings to constituent quark 
currents previously analyzed in the degenerate masses  limit \cite{EPJA-2016,PRD-2019}.
Although the relation of these quark level couplings to the 
corresponding nucleon (or baryon in general) is not fully
unambiguous for the reasons explicitated above, these 
quark level calculations and results should
lead to more fine and precise assessment of 
to what extent, why and how constituent quark model works.
Along these  lines, the absolute values of the 
resulting coupling constants are not exactly the 
most relevant or interesting point of this work, but
rather 
the effects of the nondegenerate quark masses
in each channel, i.e. neutral or charged pions
couplings to up or down quark currents, either 
pseudoscalar or axial.
This  might
 contribute to settle  a  route for analyzing hadron observable world 
with more complete grounds in low energy
     QCD. 
As pointed out in \cite{PRD-2019}
momentum dependency of 
pseudoscalar form factor within this approach
is not reasonable due to the momentum independent
 quark mass. 
Coupling constants, however, are well described.
   Since this level of calculation does not 
provide the running quark mass, the momentum 
dependency of form factors will not be
investigated in this work.
The present calculation has a dynamical character since
it corresponds to deriving an effective action for 
pions,  being mesons arranged in flavor (and spin) multiplets, 
and constituent quark currents by articulating the quark determinant
obtained from a
quark-antiquark interaction mediated by dressed gluons.
The pion-constituent quark coupling constants will be calculated 
at different kinematical points from the corresponding dynamical form factors.
This work is organized as follows.
In the next section some schematic 
explanation of the quark determinant, and its large 
quark mass expansion, in the 
presence of local meson fields and background quark currents,
surrounded by a gluon cloud, is presented.
The pseudoscalar and axial coupling constants 
are derived with the corresponding match equations.
In Section  \eqref{sec:mixings}
the possible mixings  are analyzed 
for both pseudoscalar meson states and quark currents.
For this, some mixing interactions obtained from 
flavor symmetry breaking in quark polarization \cite{PRD-2021,JPG-2022}
are also considered.
In Section \eqref{sec:numerics} numerical results for the
charged and neutral pion  pseudoscalar 
and axial coupling constants are presented
for different kinematical points: on shell and off shell (or massless) pions and constituent 
quarks.
Some strangeness degree of freedom is shown to
provide a small coupling of the strangeness quark currents 
to the neutral pion
due to the flavor mixings.
In the last Section there is a Summary with conclusions.

\section{ 
Pion couplings from the quark determinant
 }

The general procedure for the
derivation of the quark derminant
in the presence of background quark currents, surrounded
by a gluon cloud represented by  components of 
a gluon propagator,
and local  auxiliary  meson  fields displayed in
quark-antiquark flavor U(3) nonets 
 was developed   in Refs. \cite{EPJA-2016,PRD-2019,PLB-2016}.
By selecting the (chiral) scalar and pseudoscalar auxiliary fields
and by perfoming an usual
 chiral rotation,  
the scalar field  nonet is eliminated in favor of 
 non-linear parameterization of the (quasi) Goldstone boson
octet  as  fluctuations around the vacuum.
It can be written in terms of:
$U = e^{  \frac{ i  P \cdot \lambda }{F} }$
and 
$U^\dagger = e^{  \frac{ - i  P \cdot \lambda }{F} }$,
where 
$\lambda_i$ are the  flavor Gell-Mann matrices
and the pseudoscalar meson fields $P_i$ have 
canonical normalization.
The quark determinant can be written as:
\begin{eqnarray} \label{Seff-det}  
S_{eff}   &=& - i  \; Tr  \; \ln \; \left\{
 i \left[ (S_0^{f})^{-1}   +   F ( P_R U + P_L U^\dagger )
+ \sum_\phi  a_\phi  J_\phi    \right]
 \right\} ,
\end{eqnarray}
where 
$Tr$ stands for traces of all discrete internal indices 
and integration of  space-time coordinates,
$P_{R,L} = \frac{1}{2} ( 1  \pm \gamma_5)$ are the 
right-hand and left-hand
chirality projectors, 
$S_{0}^f$ is the quark propagator 
 with a 
constituent quark mass ($M_f$)
that can be 
 written
 as 
$S_0^f  = ( i \slashed{\partial} - M_f + i \epsilon )^{-1}$.
The  quark masses, $M_f$ 
for $f=u,d,s$, are   (large) constituent quark mass, usually considered to be 
generated by DChSB,  but that may
involve other mechanisms of mass generation such as 
trace anomaly as identified in Ref. \cite{EHSeff}.
The leading background dressed (singlet color) 
quark currents
  are the following:
\begin{eqnarray} \label{Rq-j}
\sum_\phi  a_\phi  J_\phi   
=
K_0 \left[  R (x-y) 
 ( J_S^i (x,y)
+ 
 i  \gamma_5 \lambda_i  J_{PS}^i (x,y) )
-
 \frac{{R}^{\mu\nu} (x-y)}{2}
 \gamma_\mu 
 \lambda_i  ( 
  \gamma_5   J_\mu^{V,i} (x,y)
+ 
 \gamma_5  J_{\mu}^{A,i} (x,y) )
\right],
\end{eqnarray}
where   
 $K_0 = 4 \alpha_g^2/9$ for 
 $\alpha_g$ encodes the running
quark-gluon coupling constant that will be  incorporated 
latter into
the effective gluon propagator.
The (non-local)  quark currents were defined respectively as
$ J_S^i  = J_S^i (x,y) =  (\bpsi \lambda^i \psi)$,
 $J_{PS}^i = J_{PS}^i (x,y) = (\bpsi i \gamma_5 \lambda^i \psi)$,
$J_{V,i}^\mu = J_{V,i}^\mu (x,y) =  (\bpsi  \gamma^\mu \lambda^i \psi)$ and
 $J_{A,i}^\mu  = J_{A,i}^\mu (x,y) = (\bpsi \gamma^\mu \gamma_5 \lambda^i \psi)$.
The background quark currents are dressed by components
   of the 
effective gluon propagator that, by assuming it is decomposed
in transverse and longitudinal terms,
can be written, in momentum space, as:
\begin{eqnarray} \label{gluonpro}
R(k) &=& 3 R_T(k) + R_L(k),
\nonumber
\\
R^{\mu\nu}(k) &=& g^{\mu\nu} (R_T (k) + R_L (k) )
+ 2 \frac{k^\mu k^\nu}{k^2 } (R_T(k) - R_L(k) ).
\end{eqnarray}
Although all the pseudoscalar nonet  have been included above,
only the components $i=1,2,3$ for the pion field  will be worked out,
 with some further attention 
to the components $i=0,8$ that undergo 
explicit flavor  meson mixing with the neutral pion
($\pi^0-\eta-\eta'$)
and, correspondingly, flavor mixing of the quark currents.

This determinant can be treated in different 
  ways.
Along the same lines of previous works, the following 
dimensionless quantites will be considered
 in the determinant:
$$
X = S_0 F (P_R U + P_L U^\dagger) ,
\;\;\;\;\;\;
Y = S_0 R (x-y) J_\phi (x.y).
$$
These quantities suggest
a   large quark (and gluon) effective mass(es) expansion
whose validity  is not directly proved due to the 
dynamical character of the involved quantities.
However,
it is interesting to note that, this same expansion for the 
NJL-model (instead of quark-antiquark interaction mediated by 
gluon exchange) provide corrections of about $20\%-30\%$ 
for the NJL coupling constant \cite{PRD-2022a,EPJA-2023} 
- values within the perturbative regime.
Also,
 it is interesting to note that, 
to provide a qualitative assessment of this expansion, 
the scalar quark current 
 can  replaced by an averaged  value in the vacuum
(rather taken its modulus as representative) that 
are the quark-antiquark scalar chiral condensate,
$| < \bar{q} q > | <   | (-250 MeV)^3 |$.
Large (quark and gluon) effective masses
 for 
relatively  low kinetic energies
(phenomenologically, it  can be considered to be below 
QCD cutoff)
yield,  for the propagators of  internal lines, $S_f \sim 1/M_f$ and 
$R \sim 1/M_g^2$ (where  $M_g$ is an effective gluon mass
that might parameterize the gluon propagator).
With such simple ad hoc consideration, 
by taking typical phenomenological values for these masses,
 $M_f \sim 350$MeV and $M_g \sim 400$ MeV, 
a typical term of this expansion can be written as
$$
Y \sim Y_{vac}  \sim \frac{ | < \bar{q} q > |  }{ M_f M_g^2 } 
\sim 
0.28,
$$
that
is not largely modified with addition 
 numerical factors and it 
 is inside the corresponding perturbative regime.
Moreover, it is worth noting that this 
expansion leads to
all  interaction
 terms 
allowed by the involved symmetries, along the lines of 
an effective field theory (EFT).
As a matter of fact, the resulting leading pion couplings  
are the ones of the Weinberg's large Nc EFT 
\cite{weinberg-2010}
with an additional scalar coupling  
(explicit symmetry breaking coupling)
as shown in \cite{EPJA-2016,PRD-2019}.
The expansion of the  whole
determinant lead to three different sectors.
Firstly, the meson sector that has bee exploited in many works
in past decades for different versions of local and non-local quark interactions,
and, specifically for the pion field,  it  yields 
Chiral Perturbation theory,
 for few works see 
\cite{meissner,GCM,ERV,chinese}.
The quark currents sector lead to
effective constituent quark interactions 
whose local limit has been exploited recently in some works
and provide corrections to models such as the NJL model.
Finally, the meson-constituent quark interactions sector
that is  exploited  
  in the local limit for pions in the present work.

\begin{figure}[ht!]
\centering
\includegraphics[width=100mm]{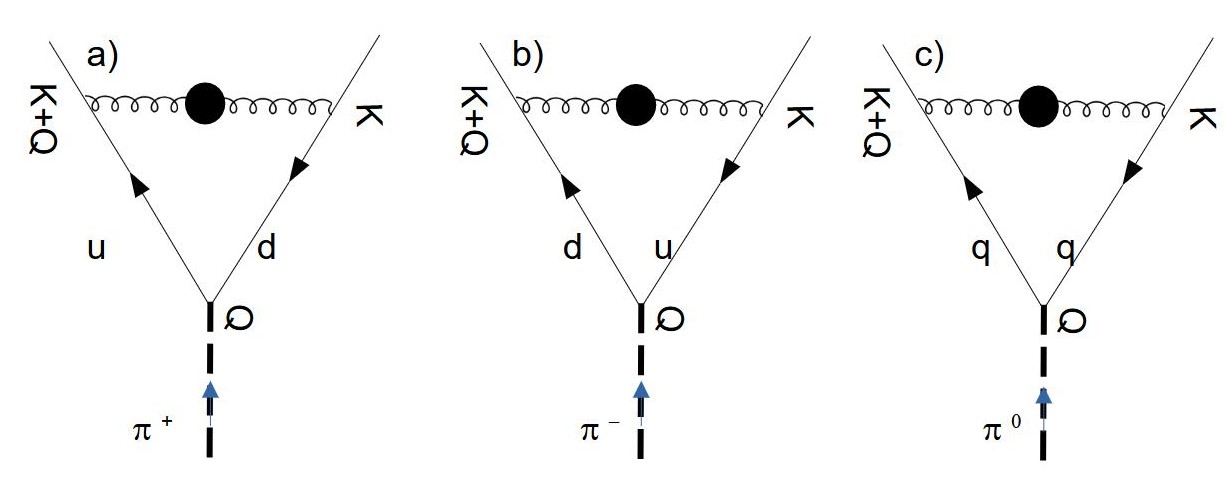}
\caption{ \label{fig:diagrams}
\small
Diagrammatic interpretation of the leading (charged and neutral) pion couplings to constituent quark currents (absorption)  at  one loop level.
Pion and quark momenta are, respectively,  $Q$ and $K$.
In Diagram (c) $q=u,d$. Wiggly lines with a dot are components of a (non-perturbative) 
gluon (effective) propagator.
 }
\end{figure}

 Eq. 
\eqref{Seff-det}
 can   be worked out, except for an irrelevant constant, as
\begin{eqnarray} \label{detXXdag}
S_{eff}^B =  - i  \; Tr  \; \ln \; \left\{ 
[ 1 + X + Y ]  \right\} = 
- \frac{i }{2} Tr \; \ln \; 
\left\{  [ 1 + X + Y ]
 [ 1 + X^* + Y^* ] \right\} .
\end{eqnarray}
The diagrammatic interpretation of the corresponding leading 
terms of the meson-quark interaction sector are shown in Fig. \eqref{fig:diagrams}
and these diagrams  - with 
one pion external leg - 
 hold for both the pseudoscalar and the axial couplings 
with different form factors/coupling constants.
Note that the flavor structure of the positive pion absorption 
coupling is different from the negative pion absorption but
it is the same as the negative pion emission.
In the expansion of the determinant, there also appear 
the scalar and vector  (two-)pion couplings \cite{PRD-2019}
that are of the same order of magnitude of the 
pseudoscalar and axial ones, although they present
a considerably more intrincate isospin structure and they 
will not be considered in this work.

\FloatBarrier


Although resulting interaction are non-local, 
by assuming large quark and gluon effective masses,
it is possible to resolve the corresponding couplings
 in the local limit. Accordingly, they can be written as:
\begin{eqnarray} \label{Gall}
I_{A} &=& G_{ij}^{ps}
 P_i J^{ps}_j + G_{ij}^A (\partial_ \mu P_i ) J^\mu_{A,j} ,
\end{eqnarray}
being that the (dimensionless) 
 coupling constants are, in fact, form factors shown below.
Alternatively,
by considering  Eq. \eqref{detXXdag},
the expansion of  the determinant 
leads to different channels of the couplings above.
The corresponding terms can be written (with B) as:
\begin{eqnarray} \label{GPSB}
I^{B} &=& G_{ij}^{ps}
 P_i^* J^{ps}_j + G_{ij}^A (\partial_ \mu P_i^* ) J^\mu_{A,j}
.
\end{eqnarray}
These coupling constants  
are in fact form factors that are   defined
according to the processes of  Fig. \eqref{fig:diagrams}.
By assuming an effective gluon propagator, that will be given in
Eq.  \eqref{gluonprop}, 
these form factors, for constant quark effective masses,
 can be defined  as:
\begin{eqnarray}  \label{Gij-ff}
G^{ps}_{i j} (K^2,Q^2, K\cdot Q) &=& 
 N_c   \; Tr_{F,D}
\int \frac{ d^4 k}{(2\pi)^4} 
S_0^{f_1} (k+K) i \gamma_5 \lambda_i S_0^{f_2} (k+K+Q) \lambda_j 
i \gamma_5 R(-k), 
\\
G_{ij}^A  (K^2,Q^2, K\cdot Q)  g^{\mu\nu} &=&
    N_c  F  \; M_{f_2} \;  Tr_{F,D}
\int \frac{ d^4 k}{ (2\pi)^4}
\tilde{S}_0^{f_1} (k+K) \gamma^\mu \gamma_5 \lambda_i 
\tilde{S}_0^{f_2} (k+K+Q) \lambda_j 
\gamma^\nu \gamma_5  R (-k), 
\nonumber
\end{eqnarray}
where $Tr_{F,D}$ is the trace in flavor and Dirac indices,
$\tilde{S}_0^f   (k) = ( k^2 - M_f^2 + i\epsilon )^{-1}$.
Although only the states $i,j = 1,2,3$ 
will be worked out, some other 
states are  related by mixings, $i,j=0,8$.
For the definitions above, 
  in the case of nondegenerate quark mass 
 $G_{f_1 \neq f_2} = G_{f_2 \neq f_1}$.
The difference is, however, very small being proportional to 
the up and down quark mass differences in the isospin subgroup.
Due to U(1) electromagnetic invariance and 
CP, one has $G_{i j} = G_{j i}$ and $G_{11}=G_{22}$.
Some
  complementary  definitions
that will be useful to work out mixings 
are presented in the Appendix \eqref{sec:appB}.
Positive and negative pions  
present exactly the same form factors
since $G_{11}$ and $G_{22}$ are given by Eq. \eqref{G11}.
In fact, it involves the 
average of positive  or negative pion absorption and emission 
processes.
Note that the equations for the axial and pseudoscalar 
pion couplings are independent of each other in the 
approach considered in the present work, and 
none of them is calculated by considering the 
GTR which, nonetheless, is  satisfied quite well
when comparing numerical results.
It is also interesting to emphasize that, as a consequence
of the standard methods applied for the 
derivation of the quark determinant,
  pion couples directly to the pseudoscalar and axial 
quark currents.
It is worth to emphasize that
the  above form factors were obtained in a dynamical way and 
not from the QCD vertex function as calculated diagrammatically or in a Schwinger-Dyson equations framework.
Therefore
 coupling constants do not 
emerge 
 as the residue of the pole 
 from the corresponding Bethe-Salpeter equation (BSE)
\cite{arriola,HORN-ROB}.
The relation between the present approach
and the standard one based in  BSE
has not yet been settled.

\section{ Quark currents, meson states and mixings}
\label{sec:mixings}

By writing  interactions \eqref{GPSB} explicitely in terms of 
charge eigenstates, 
for  the  pseudoscalar  coupling and corresponding pion form factors,
it yields:
\begin{eqnarray} \label{GPSBud}
  I_{ps}^B
 &=& 
 G_{11}^{ps} 
 ( \pi^+ \bar{d} i\gamma_5  u + \pi^- \bar{u}  i\gamma_5  d )
+ 
 G_{33}^{ps}  \pi_3 ( \bar{u}  i\gamma_5 u - \bar{d}  i\gamma_5 d ),
\end{eqnarray}
And similarly for the axial couplings.
The charged pion couplings can be written as:
\begin{eqnarray}
I_{\pi^\pm}^{ps} = ( G_{ud}^{ps} + G_{du}^{ps} )
 ( \pi^+  \; \bar{d} i \gamma_5 u 
+ \pi^-  \; \bar{u} i \gamma_5 d ),
\end{eqnarray}
where $G_{f_1f_2}$ are defined in Eqs. \eqref{G11}-\eqref{G88}.
Note that these couplings correspond to an average
of  emitted and absorbed positive or negative pion.
The separate coupling for emission or absorbed pion
are equal in the limit of degenerate quark masses.

\subsection{ Mixings}

The complete pseudoscalar interaction
 terms  can be written, for the pion, i.e.  $i,j=1,2,3$ and
the corresponding mixing terms, $(i,j)=(0,3)$ or 
$(i,j)=(3,8)$, as:
\begin{eqnarray} \label{IB-Gij2}
I_{ps}^{mix-ij} \equiv G_{ij}^{ps}  P_i J_j^{ps}
=  G_{11}^{ps} P_1 J_1^{ps}  + G_{22}^{ps} P_2 J_2^{ps} 
+ G_{33}^{ps} P_3 J_3^{ps} 
+ G_{38}^{ps} P_3 J_8^{ps}  
+ 
G_{30}^{ps} P_3 J_0^{ps} ,
\end{eqnarray}
Being the flavor structure of the  axial couplings analogous.
For the neutral pion, 
if the mixing terms in Eq. \eqref{IB-Gij2} are  not 
taken into account,  the neutral pion coupling
  becomes
the following one:
\begin{eqnarray} \label{G33-nomixing} 
I_{\pi^0}^{no mixing} =  (G_{uu}^{ps} + G_{dd}^{ps} )
 P_3 \; \left(  \bar{u} i \gamma_5 u 
-
 \bar{d}  i \gamma_5 d \right) = 
G_{ps}^{\pi^0} (0) \; P_3 \; \bar{\psi} \lambda_3 i\gamma_5 \psi.
\end{eqnarray}
From this equation the following coupling constant,
without mixing, can be defined
$G_{ps}^{\pi^0}(0)$), and similarly to the axial coupling constant.
If the  explicit mixing terms between
 the neutral (either meson or current)
 states $i,j = 0, 3, 8$ are considered, 
 by writing explicitly the interactions,
 the charged and neutral pion couplings can be rewritten as:
\begin{eqnarray} \label{Guudd-mix1}
I_{\pi^0}^{mixing} &=&
G_{uu}^{ps} P_3 \; \bar{u} i \gamma_5 u 
-
  G_{dd}^{ps}  P_3 \; \bar{d}  i \gamma_5 d .
\end{eqnarray}
Therefore the neutral pion couples differently to 
up or down quark currents.
However, the neutral pion coupling to
a constituent 
quark current might  be, still,  interpreted as an average
of the up and down couplings 
so that Eq. \eqref{G33-nomixing}
 seems reasonable in the presence of these mixings,
i..e
\begin{eqnarray} \label{Gpspi0mix}
I_{\pi^0}^{mixing} \sim G^{ps}_{\pi^0}(ij) P_3 \; 
(\bar{\psi} i\gamma_5 \lambda_3 \psi).
\end{eqnarray}

Another mixing term can be taken into account.
 For instance, there might appear
neutral pions from to the oscillations  of $\eta,\eta'$ 
 due to meson mixings. 
There are two different possible ways meson mixing
can show up. 
First in processes that lead to 
pions, etas and etaprimes for which contaminations
from $\eta,\eta'$ oscillations can lead to 
contributions for the neutral pion coupling to constituent quarks/nucleon.
Second, 
in processes in which the original neutral pion oscillates to
$\eta$ or $\eta'$ that can couple to quarks/nucleons.
Consider the following parameterization of these meson mixings
(rotations):
\begin{eqnarray} \label{mesonmix}
P_0 &=& 
 \bar{G}_{80}\eta + \bar{G}_{03} \pi^0 + a_0 \eta'
\nonumber
\\
P_8 &=& \bar{G}_{08} \eta' + \bar{G}_{83} \pi^0 
+ a_8 \eta, 
\nonumber
\\
P_3 &=& a_3 \pi^0 +  \bar{G}_{38} \eta +  
\bar{G}_{30} \eta' ,
\end{eqnarray}
where the pion receives the smallest  mixings and then
$a_3 < 1$ with $a_3   \sim 1$.
These equations have been 
inspired in usual parameterizations \cite{mixing1,mixing2}. 
Only two mixing terms will be needed 
below,
$\eta-\pi^0$ or $\eta'-\pi^0$  
by means of  $\bar{G}_{03}$ and $\bar{G}_{38}$.
Furthermore, two types of interactions will be analyzed, the pseudoscalar and axial, so that all the coupling constants 
and mixing parameters will be labeled with  an extra index
$\xi = A,  ps$ below.
As a consequence, 
the following neutral pion coupling to quark currents, 
$J_u^\xi, J_d^\xi, J_s^\xi$,  
 in both  axial and pseudoscalar channels,
 are obtained:
\begin{eqnarray}  \label{Gpi0mix}   
L_{\pi^0-J}^\xi &=&
(\partial_\xi \pi^0) a_3 \left[
J_u^\xi \left( G_{uu}^\xi + \frac{ \bar{G}_{38}^\xi G_{uu}^\xi }{a_3}
+ \frac{ 2 \bar{G}_{03}^\xi G_{uu}^\xi }{ \sqrt{3} a_3 }
\right)
- J_d^\xi \left( G_{dd}^\xi 
- \frac{ \bar{G}_{38}^\xi  G_{dd}^\xi  }{a_3}
- \frac{ 2 \bar{G}_{03}^\xi  G_{dd}^\xi  }{ \sqrt{3} a_3 }
\right)
\right.
\nonumber
\\
&+& \left.  \sqrt{2} J_s^\xi G_{ss}^\xi \left( 
\bar{G}_{38}^\xi  - \frac{ \bar{G}_{03}^\xi  }{\sqrt{3}} \right)
\right]
\nonumber
\\
&=& 
(\partial_\xi \pi^0)  a_3  \left[ 
  G_{mix,uu}^\xi J_u^\xi
-  G_{mix,dd}^\xi J_d^\xi
+   G_{mix,ss}^\xi J_s^\xi
\right],
\end{eqnarray}
where:
\begin{eqnarray}  \label{definitions}
\mbox{for} \;\;\;\;
\xi = ps &:& (\partial_\xi \pi^0)  \to \pi^0,
\;\;\;\;\;\;\;\;  J^\xi = J^{ps},
\nonumber
\\
\mbox{for} \;\;\;\;\;
\xi = A  
&:&   (\partial_\xi \pi^0)  \to  \partial_\mu \pi^0,
\;\;\;\;\;
J^\xi = J^\mu_A.
\end{eqnarray}
Note the last term 
 is a neutral pion coupling to the strange quark (axial or pseudoscalar) current.
In the last  line of Eq. \eqref{Gpi0mix},
 the neutral pion coupling to the different
quark currents incorporated the mixings that lead to the 
neutral pion coupling to the strange
quark current, $G^{\xi}_{mix,ss}$.
The neutral pion couplings
are corrected differently for $\bar{u}u$ and $\bar{d}d$ states,
i.e. the corresponding corrections  have  opposite signs.
By resorting to an  analogy with the
 decomposition for Eq. \eqref{G33-nomixing} 
that yielded Eq. \eqref{Guudd-mix1}, the 
following neutral pion coupling to constituent quarks
will be  defined by using the definitions \eqref{definitions}:
\begin{eqnarray} \label{Gpi0mixJ2}
I_{\pi^0,mix}^{mix} &=& 
(\partial_\xi \pi^0)  a_3 (G_{mix,uu}^\xi (ij)
 + G_{mix,dd}^\xi (ij)  )
J^\xi_3.
\end{eqnarray}

The  meson  mixing parameters  \eqref{mesonmix}
will be extracted from a simple association of 
 meson mixing interactions $G_{i \neq j}$ that 
lead to diagonal meson states by starting from
flavor eigenstates 
by considering separate diagonalization/procedure for 
each of the mixing interactions
\cite{PRD-2021,JPG-2022}.
This, nearly {\it ad hoc}, prescription yields:
\begin{eqnarray} \label{Gbar}
g_{03}^\xi  \equiv \bar{G}_{03}^\xi = \frac{ G_{03}^\xi  }{2 ( G_{00}^\xi +  G_{33}^\xi ) },
\;\;\;\;\;
g_{38}^\xi  \equiv  
\bar{G}_{38}^\xi = \frac{ G_{38}^\xi }{2 ( G_{33}^\xi +
 G_{88}^\xi  ) }.
\end{eqnarray}
If these mixings were simply
$\bar{G}_{03}^\xi = G_{03}^\xi$ 
and $\bar{G}_{38}^\xi = G_{38}^\xi$
the pion coupling to strange quark current would 
disappear, i.e. $G_{mix,ss}^\xi = 0$.

\subsection{ Renormalization condition}

To provide numerical results, the folllowing 
renormalization condition will be considered
by fixing the pseudoscalar charged pion
 coupling constant with a value
based in  phenomenology \cite{GPS}
 (off shell pion and  on shell quark):
\begin{eqnarray} \label{recondition}
G^{ps}_{\pi^\pm}
 (K^2=M^2, Q^2=0) = 13.
\end{eqnarray}
It turns out that, for the sets of parameters 
considered in this work,
results will provide  nearly
$g_A (M^2, 0)^{\pi^\pm}
 \sim  1$ without further 
conditions.
Conversely, in other works \cite{PRD-2019,PRD-2018a},
 an alternative condition was considered, i.e.  to 
fix  either $g_A = 1$ or $g_v=12$ (vector meson coupling 
to constituent quarks) for the CQM.
A different condition could be imposed by assuming
each quark of the nucleon  to be responsible for 
part of the couplings to mesons
along other investigations \cite{rijken}, such 
that $G_{ps} \sim 13/3$.
However, for the condition above,
pion-nucleon coupling constants should arise 
from an average of pion-constituent quark coupling constants
instead as a sum.
This might be considered 
 in the  Goldberber-Treiman (GT)-relation if  the nucleon mass
is considered instead of the constituent quark mass ($M_q$).
Otherwise, with a constituent quark mass, GT-relation should be written 
as 
$\frac{G^{ps} }{g^A}  = \frac{M_q}{f_\pi} ( 1 + \Delta_{GT})$ being 
$\Delta_{GT}$ not small with respect to 1.

%
%
%

\section{ Numerical Results}
 \label{sec:numerics}

The following  longitudinal (confining) effective gluon propagator,
inspired in \cite{cornwall} and tested in different works \cite{PRD-2018a,PRD-2019},
has been
adopted:
\begin{eqnarray} 
\label{gluonprop}
D^{\mu\nu}  (k) &=&   \frac{ K_F g^{\mu\nu}  }{ (k^2 - M_g^2)^2 },
\end{eqnarray}
where the strong running coupling constant ($\alpha_g^2$) 
 has been incorporated implicitely in $K_F$.
It will be considered $M_g = 500 $MeV,  
 and the normalization $K_F$  is determined by imposing  the renormalization condition \eqref{recondition}.
It has been shown, in Refs. \cite{PRD-2018a,PRD-2019},
 to produce similar  numerical values and behavior
 to a different  gluon propagator extracted from 
Schwinger Dyson equations \cite{SDE}
for
 spacelike meson-constituent quark form factors  

%
%
%
%
%
%

Given that there is an intrinsic ambiguity in defining values for
the quark effective masses, 
 results will be shown for different choices
 of 
$M_u, M_d$ (and also $M_s$).
These masses, for the chosen gluon effective propagator,
determine  $G_{ij}^\xi$ (both $G^{ps}$ and $G^{A}$)
 numerically, and the different
 sets of parameters 
(s.o.p.) are shown in Table \eqref{table:SOP}.
In the present work, we consider  quark 
effective masses - 
sets of parameters $A_1$ and $A_2$ - 
 obtained from a (flavor-dependent) NJL model
as presented in \cite{PRD-2021}.
In NJL-type models
 there is   need to fix chosen
observables/quantities to be mass independent
and this choice may bring differences in the results,
for example in Refs. \cite{CQM-baryon},
being the neutral and charged pion masses  chosen to fit the model parameters.
In the present calculation however, the renormalization condition
presented in Eq. \eqref{recondition} is a further "constraint"
for  the resulting effective action and numerical results.
The third s.o.p., $A_3$, is based in values  often  found in the literature.

\begin{table}[ht]
\caption{
\small Sets of parameters for
quark effective  masses. 
} 
\centering  
\begin{tabular}{c c c c c c c c }  
\hline\hline  
set  of  & $M_u$ &   $M_d$ &  $M_s$ 
\\
parameters &  MeV  &       MeV     &    MeV  
 \\
\hline
\\ [0.5ex]
$A_1$  
&  389  & 399  & 600 
\\ [0.5ex]
$A_2$ &  307  & 319 & 349 
\\ [0.5ex]
$A_3$ & 340  & 350  & 550 
\\
\hline
\\
$m_{\pi^0}$= 134.98 ,& 
$m_{\pi^\pm}$= & 139.57 & (MeV)
\\[1ex] 
\hline 
\end{tabular}
\label{table:SOP} 
\end{table}
\FloatBarrier

The mixing parameters defined in Eq. \eqref{mesonmix}
 were calculated at the same kinematical point ($K^2=M_f^2,Q^2=0$)  for
each of the s.o.p. 
The relative difference 
from one kinematical point to another is much smaller.
They are given by
\begin{eqnarray}
A_1  && \bar{G}_{03}^{ps}  = 6.3 \times 10^{-3},
 \;\;\;\;\; 
 \bar{G}_{38}^{ps}  =  5.9 \times 10^{-3},
 \;\;\;\;\; 
 \bar{G}_{03}^{A}  = -1.6 \times 10^{-3} ,
 \;\;\;\;\; \bar{G}_{38}^{A}  =  -1.1 \times 10^{-3}
,
\nonumber
\\
A_2  && \bar{G}_{03}^{ps}  = 3.9 \times 10^{-3} ,
 \;\;\;\;\; 
 \bar{G}_{38}^{ps}  = 2.8 \times 10^{-3}
 \;\;\;\;\; 
 \bar{G}_{03}^{A}  = -2.7  \times 10^{-3} ,
 \;\;\;\;\;  \bar{G}_{38}^{A}  =  -1.9 \times 10^{-3}
,
\nonumber
\\
A_3  && \bar{G}_{03}^{ps}  = 4.5 \times 10^{-3} ,
 \;\;\;\;\; 
 \bar{G}_{38}^{ps}  = 3.8 \times 10^{-3}
 \;\;\;\;\; 
 \bar{G}_{03}^{A}  = -1.9  \times 10^{-3} ,
 \;\;\;\;\; 
\bar{G}_{38}^{A}  =  -1.3 \times 10^{-3}
,
\end{eqnarray}


In Table \eqref{table:Gpstable} values of
pseudoscalar coupling constants
$G_{ps}$, defined in different kinematical points
(off shell or on shell
constituent quark and pion)
 are shown for the  different sets of parameters 
presented in Table \eqref{table:SOP}. 
The following kinematical points have been considered:
\begin{eqnarray}
&  S_1:  (K^2 , Q^2) = (M_f^2, 0),  &  \;\;\;
S_2:  (K^2 , Q^2) = (M_f^2, M_\pi^2), 
\nonumber
\\
& S_3:  (K^2 , Q^2) =(0, 0), & \;\;\;
S_4: (K^2 , Q^2) = (0, M_\pi^2).
\end{eqnarray}
Besides the charged pion coupling constant,
the neutral pion coupling constant is exhibitted    
with and without  flavor states mixings, 
respectively
$G^{ps}_{ \pi^0}(0)$ and $G^{ps}_{ \pi^0}(ij)$ from 
Eqs. \eqref{G33-nomixing} and \eqref{Gpi0mixJ2}.
As shown above the neutral pion coupling 
can be decomposed in the couplings of states
$\bar{u}u$ and $\bar{d}d$ states, that might also 
receive contributions from flavor state mixings
- and these are shown in the middle part of this
Table.
In the last two columns  the pseudoscalar 
coupling constant of the neutral
pion coupling to strange quark current $G_{mix,ss}^{ps}$
is exhibitted  with a 
parameter $x_{ps}^{m}$ that will be defined below.

Results for s.o.p. $A_2$ ($A_1$), with the lower (larger) values
of quark masses,  present the smallest (largest) variation of values
for the different kinetical points.
The charged pion coupling constants
 are smaller than the neutral  pion ones 
 except for  the case of the fourth kinematical point:
off shell quarks and on shell pions $(K^2=0,Q^2=M_\pi^2)$.
The neutral pion coupling to the $\bar{u}u$ quark current
has larger strength than the coupling to $\bar{d}d$
for all kinematical points, and  for both
cases, without and with mixings, $G^{ps}_{uu}(0)$ and $G^{ps}_{uu}(ij)$ 
respectively.

Recent analysis  
of the isospin dependence of the
pion-nucleon pseudoscalar coupling constant 
have been done
  with 
different approaches \cite{granada,bochum}.
For
coupling constants, usually denoted by
 $f_{\pi^a NN'}$ for $a=0,\pm$ and $N,N'=n,p$,
results  
exhibit  basically the same
trend
with small numerical  differences between charged and neutral 
pion couplings to nucleons. 
For the sake of comparison of the behavior,
some  values  are reproduced   in Table \eqref{tab:valuesGps}.
The first one  is the value extracted  from Ref. \cite{bochum},
the second and third lines correspond to  representative 
values   from
Ref. \cite{granada}.
In spite of the differences in absolute values,  
 when comparing to Table
\eqref{table:Gpstable},
the analysis will be focused on the relative values.
The
coupling constant of  neutral pion  to neutrons is larger than 
the  coupling constant  to protons 
and 
 the one for charged pion in both approaches.
The hierarchy of values is nearly the same in all these 
sets of coupling constants and they are in agreement with 
general results obtained from NJL-type models in Ref.
\cite{ArriolaAmaroPerez}.
In the following, a model is presented to allow for an analysis of 
the possible relevance of the present work for the 
pion coupling to nucleons.

\begin{table}[ht]
\caption{
\small  
Neutral and charged pion-nucleon pseudoscalar coupling constants extracted
from different recent works \cite{granada,bochum}.
 }
\centering  
\begin{tabular}{| c | c c c  | }  
\hline\hline  
                  & $G_{\pi^0nn}$ &   $G_{\pi^0pp}$ &  $G_{\pi^\pm np}$ 
\\
\hline
$1$- \cite{bochum} & 
13.397 &   13.227 & 13.226
\\ [0.5ex]
$2$- \cite{granada} &  13.415  & 13.148  & 13.252  
\\ [0.5ex]
$3$- \cite{granada} & 13.321  & 13.226  & 13.226
\\[1ex] 
\hline 
\end{tabular}
\label{tab:valuesGps} 
\end{table}
\FloatBarrier

According to most versions of the quark model,
 proton and neutron 
couple to mesons by means of quarks in an addictive 
way \cite{LeYaouanc1973,rijken}.
Along this idea,
the pion coupling to nucleons
might also arise in a sort of  
average for the pion coupling to corresponding quark currents,
$p \sim uud$ and $ n \sim ddu$.
Due to 
 the
choice of the 
renormalization condition 
($G^{ps}_{\pi^\pm}  =13$),
the resulting neutral pion coupling to nucleons, 
as a first simplified approach, might be considered
as a simple average:
\begin{eqnarray} \label{CQM-nucleon}
g_{\pi^0 nn} \sim \frac{2 G_{\pi^0 dd} + G_{\pi^0 uu}}{3} ,
\;\;\;\;\;\;\;\;
g_{\pi^0 pp} \sim \frac{ 2 G_{\pi^0 uu} + G_{\pi^0 dd}}{3} .
\end{eqnarray}
The behavior shown in 
the Table \eqref{table:Gpstable}  is partially consistent with the results
from
 Refs.  \cite{granada,bochum}. 
There are, however, few different behaviors.
From Table \eqref{tab:valuesGps}
it is seen that 
 $g_{\pi^0 pp} \sim  g_{\pi^\pm np}$ 
(or 
 $g_{\pi^\pm np} >  g_{\pi^0nn}$  from 2-\cite{granada}).
This is consistent with results from Table \eqref{table:Gpstable}
being that $g_{\pi^\pm np} >  g_{\pi^0nn}$ is found for the kinematical 
point $S_4$.
However, 
for the results shown in the Table  \eqref{table:Gpstable} it
would be   expected that
$g_{\pi^0 nn} <  g_{\pi^0 pp}$ and this is not compatible
with the results shown in Table \eqref{tab:valuesGps}.
 
It is shown below that mixings can 
modify the relative  behavior of $G_{uu}^{ps}(ij)$ and 
$G_{dd}^{ps}(ij)$,
and then  $g_{\pi^0nn}$ and $g_{\pi^0pp}$.
By  varying the strength of the mixing interactions by an arbitrary
 constant uniform 
factor $x^m_{ps}$, 
the pseudoscalar  coupling constants $G^{ps}_{uu}$ and $G^{ps}_{dd}$ 
can flip their hierarchy
(suitable for making $g_{\pi^0 pp} < g_{\pi^0 nn}$)
 if $x_{ps}^m < 0$.
Therefore  mixing parameters were redefined by:
\begin{eqnarray} \label{xmps}
\bar{G}^{ps}_{i \neq j} \to x^m_{ps}
 \times  \bar{G}^{ps}_{i \neq j}.
\end{eqnarray}
This redefinition of the mixing parameters can  
include further mixing interactions, not only due to 
isospin/flavor symmetry breaking, i.e. 
they might include 
't Hooft interactions.
The resulting values for $x_{ps}^m$ for which the 
coupling constants $G^{ps}_{uu}$ become smaller than
 $G^{ps}_{dd}$
are in the last column of the Table \eqref{table:Gpstable}.
However, with this {\it ad hoc} change in the mixing parameters to 
make quark interactions compatible with neutral pion coupling to nucleons
the hierarchy of neutral and charged pion couplings
may be inverted unless
a redefinition of the charged pion coupling constant
is done.
  

In  Table \eqref{table:GAtable}
values of the axial $G_{A}$
 coupling constant
 are shown for the same cases 
of  Table \eqref{table:Gpstable}.
The behavior for $G_A$ at different kinematical points
 are different from  the results of the $G_{ps}$, 
being the charged pion coupling constants slightly larger than
the neutral pion ones.
Although 
the variation  of the axial coupling constant is smaller than the 
case of the pseudoscalar coupling constant of the previous Table, 
the relative deviations are nearly of the same order of magnitude.
 As discussed above, 
the strength of the  pseudoscalar coupling constant 
is fixed with the 
renormalization condition \eqref{recondition}
and this settles the values shown in the Table
 $G_A(K^2, Q^2)$.
The s.o.p.   $A_2$ (lower quark effective masses)
presents the closer value  of $G_A$ 
to the 
expected value for the CQM, $G_A = 1$  \cite{weinberg,weinberg-2010}.
However, the author is not aware of previous 
investigations of  the contribution 
of the isospin breaking for the pion axial coupling constant
in the literature. 
In the last column of Fig. \eqref{table:GAtable}
the coupling constant for the axial pion - strange quark current 
is presented.
The strength of this {\it strange} axial pion coupling constant
is relatively smaller than the pseudoscalar one.
The axial $G^{uu}_{A}$ are already weaker than $G^{dd}_{A}$
and the mixing interactions increase this difference.


%

\begin{table}[ht]
\caption{
\small 
Pseudoscalar pion coupling constants for charged and 
neutral pions at different kinematical points,
without ($G^{ps}_{\pi^0} (0)$) \eqref{G33-nomixing}
 and with mixing interactions $G_{i\neq j}$ ($G^{ps}_{\pi^0} (ij)$)
\eqref{Gpi0mixJ2}.
Values of pseudoscalar coupling constants
for the $\pi^0$ with states $\bar{u}u$ and $\bar{d}d$
(shortened notation  $G^{ps}_{uu}$ and $G^{ps}_{dd}$)
  also shown without $(0)$  and with $(ij)$
mixing interactions.
In the two last columns, 
the pseudoscalar coupling constant of neutral pion -strange quark current interaction
and the parameter used  to vary mixing interactions $x_{ps}^m$ 
defined in Eq. \eqref{xmps}.
} 
\centering  
\small
\begin{tabular}{ | c  |  c c c  | c  c c  c |   c | c  | } 
\hline\hline  
s.o.p. & $G^{ps}_{\pi^\pm}$ &  $G^{ps}_{\pi^0}$  (0)  &
  $G^{ps}_{\pi^0}$ ($ij$) & $G^{ps}_{uu}$  (0)
&  $G^{ps}_{uu}$  ($ij$) & $G^{ps}_{dd}$ (0)   &  $G^{ps}_{dd}$  ($ij$)
& $G^{ps,ss}_{mix}$ & $x^m_{ps}$ 
\\
\hline \hline
$A_1$ & & & & &   & & & &
\\
\hline
$S_1$ & 
13.000(1)   &  13.001(2)  & 13.007(2) & 13.471(1)   & 13.648(1)  & 12.531(1) & 12.367(1)
&  0.05(1) & -1.7(2)
\\ 
\hline 
  $S_2$ & 
10.446(1) & 10.546(2) & 10.553(2)  & 11.098(1) & 11.244(1)  & 9.994(1) & 9.863(1) 
& 0.37(1) & -1.2(2)
\\ 
\hline 
 $S_3$  & 
17.870(1)   & 17.871(2)  & 17.875(2) & 18.172(1) & 18.411(1) & 17.569(1) & 17.338(1) 
& - 0.01(1) & -2.3(2)
\\ 
\hline 
$S_4$ & 
18.106(1) &  18.091(2) & 18.095(2) &  18.399(1) &  18.642(1) &  17.782(1)  &  17.545(1)
&-0.01(1)  & -2.3(2)
\\ 
\hline \hline
$A_2$ & & & & &   & & & &
\\ 
\hline
$S_1$ & 
13.000(1)  & 13.000(2) & 13.003(2)  & 13.363(1) & 13.463(1) & 12.637(1) & 12.543(1)
& -0.02(1) & -2.6(2)
\\ 
\hline 
$S_2$ & 
 12.263(1)   & 12.284(2)  & 12.287(2)  & 12.708(1) & 12.802(1) & 11.861(1) & 11.773(1)
& -0.02(1)  & -2.5(2)
\\ 
\hline 
$S_3$ & 
 14.550(1)   & 15.550(2) & 14.552(2)  & 14.818(1) & 14.928(1) & 14.283(1) & 14.177(1)
& -0.01(1) & -2.6(2)
\\ 
\hline 
$S_4$ & 
14.808(1) &  14.792(2)  &  14.794(2) & 15.069(1) &  15.181(1) &  14.514(1) & 14.406(1)
& -0.01(1)  & -2.2(2)
\\ 
\hline  \hline
$A_3$  & & & & &   & & & &
\\
\hline 
$S_1$ & 
13.000(1) & 13.000(2)  & 13.005(2) & 13.352(1) & 13.537(1)  & 12.649(1) & 12.474(1)
& - 0.004(5) & -2.0(2)
\\ 
\hline 
$S_2$ &
 11.735(1)  & 11.779(2)   & 11.783(2)  & 12.191(1) & 12.301(1)  & 11.368(1)  & 11.266(1) 
& 0.04(1) & -1.8(2)
\\ 
\hline 
$S_3$
& 15.458(1)   & 15.459(2)   & 15.461(2)  & 15.704(1)  &  15.846(1) & 15.213(1) & 15.076(1) 
& -0.01(1) & -2.3(2)
\\ 
\hline 
$S_4$ & 
15.702(1) & 15.686(2) & 15.689(2) & 15.940(1) & 16.083(1)  & 15.433(1) & 15.294(1)
&  -0.01(1) & -2.3(2)
\\ 
\hline \hline
\end{tabular}
\label{table:Gpstable}  
\end{table}
\FloatBarrier

\begin{table}[ht]
\caption{
\small 
Axial pion coupling constants for charged and neutral pions
at different kinematical points, 
without ($G^A_{\pi^0} (0)$)
 and with  ($G^{A}_{\pi^0} (ij)$) mixing interactions $G_{i\neq j}$,
similarly to 
Eqs. \eqref{G33-nomixing}
 and
\eqref{Gpi0mixJ2}.
Values  presented  for the 
 the states $\bar{u}u$ and $\bar{d}d$ states without $(0)$ and with 
$(ij)$
mixing interactions.
In the last column
the axial  coupling constant of neutral pion -strange quark current. 
} 
\centering  
\begin{tabular}{|| c  ||  c c c  | c  c c  c | c  ||} 
\hline\hline   
 s.o.p. & $G^{A}_{\pi^\pm}$ &  $G^{A}_{\pi^0}$  (0)  &  $G^{A}_{\pi^0}$ ($ij$) & $G^{A}_{uu}$  (0)
&  $G^{A}_{uu}$  ($ij$) & $G^{A}_{dd}$ (0)   &  $G^{A}_{dd}$  ($ij$)
& $G^{A}_{mix,ss}$ 
\\
\hline \hline
  $A_1$ & & & & &   & & &  
\\
\hline
$S_1$
&  1.603(1)   & 1.603(2)   &  1.603(2)  & 1.600(1)   & 1.595(1)  & 1.605(1)  & 1.610(1)  
& 0.001(1) 
\\ 
\hline 
$S_2$
 & 1.889(1)  
 & 1.875(2)  & 1.875(2)    & 1.872(1)   &  1.866(1)  & 1.878(1)   & 1.8834(1)  
&  0.001(1) 
\\ 
\hline 
$S_3$ 
 & 1.194(1)   & 1.194(2)    &  1.194(2)   & 1.197(1)   & 1.194(1)  & 1.190(1)   &  1.194(1) 
& - 0.001(1) 
\\
\hline
$S_4$ & 
1.223(1)  &  1.221(2)  & 1.221(2)   & 1.225(1)  &  1.221(1)   & 1.217(1)  &  1.221(1) 
& - 0.001(1) 
\\ 
\hline \hline
$A_2$ & & & & &   & & &  
\\
\hline
$S_1$ &
 0.974(1) 
 & 0.974(2)  & 0.974(2)  & 0.971(1)   & 0.966(1)   & 0.976(1)  & 0.982(1)  
&   0.001(1) 
\\ 
\hline 
$S_2$  &
 1.144(1)  
 & 1.136(2)   & 1.136(2)   & 1.132(1)   & 1.126(1)   & 1.139(1)  & 1.145(1) 
& 0.001(1) 
\\ 
\hline 
$S_3$  &
 0.7801(1)
 & 0.780(2)   & 0.780(2)   & 0.782(1)   & 0.778(1)   & 0.778(1)  & 0.782(1)   
& -0.001(1) 
\\
\hline
$S_4$ & 
0.806(1)   & 0.804(2)  &  0.804(2)   & 0.807(1)   & 0.803(1)   & 0.802(1)  & 0.806(1) 
& -0.001(1) 
\\ 
\hline \hline
$A_3$ & & & & &   & & &  
\\
\hline
$S_1$ &
 1.160(1)   &  1.160(2)   &  1.160(2)  &  1.158(1)   &  1.154(1)  & 1.163(1)  & 1.167(1) 
& 0.001(1) 
\\ 
\hline 
$S_2$  &
 1.365(1) 
 &  1.355(2)  &  1.355(2)   & 1.352(1)   & 1.347(1)   & 1.358(1)  & 1.363(1)  
& 0.001(1) 
\\ 
\hline 
$S_3$
 & 0.903(1)  & 0.903(2)   &  0.903(2)   & 0.905(1)   & 0.902(1)   & 0.901(1)  & 0.904(1) 
& -0.001(1) 
\\
\hline
$S_4$ & 
0.930(1)  & 0.928(2)  & 0.928(2)  & 0.931(1)  & 0.927(1)  
& 0.925(1)   & 0.928(1) 
& -0.001(1) 
\\ 
\hline \hline
\end{tabular}
\label{table:GAtable}  
\end{table}
\FloatBarrier

%

%

\begin{figure}[ht!]
\centering
\includegraphics[width=100mm]{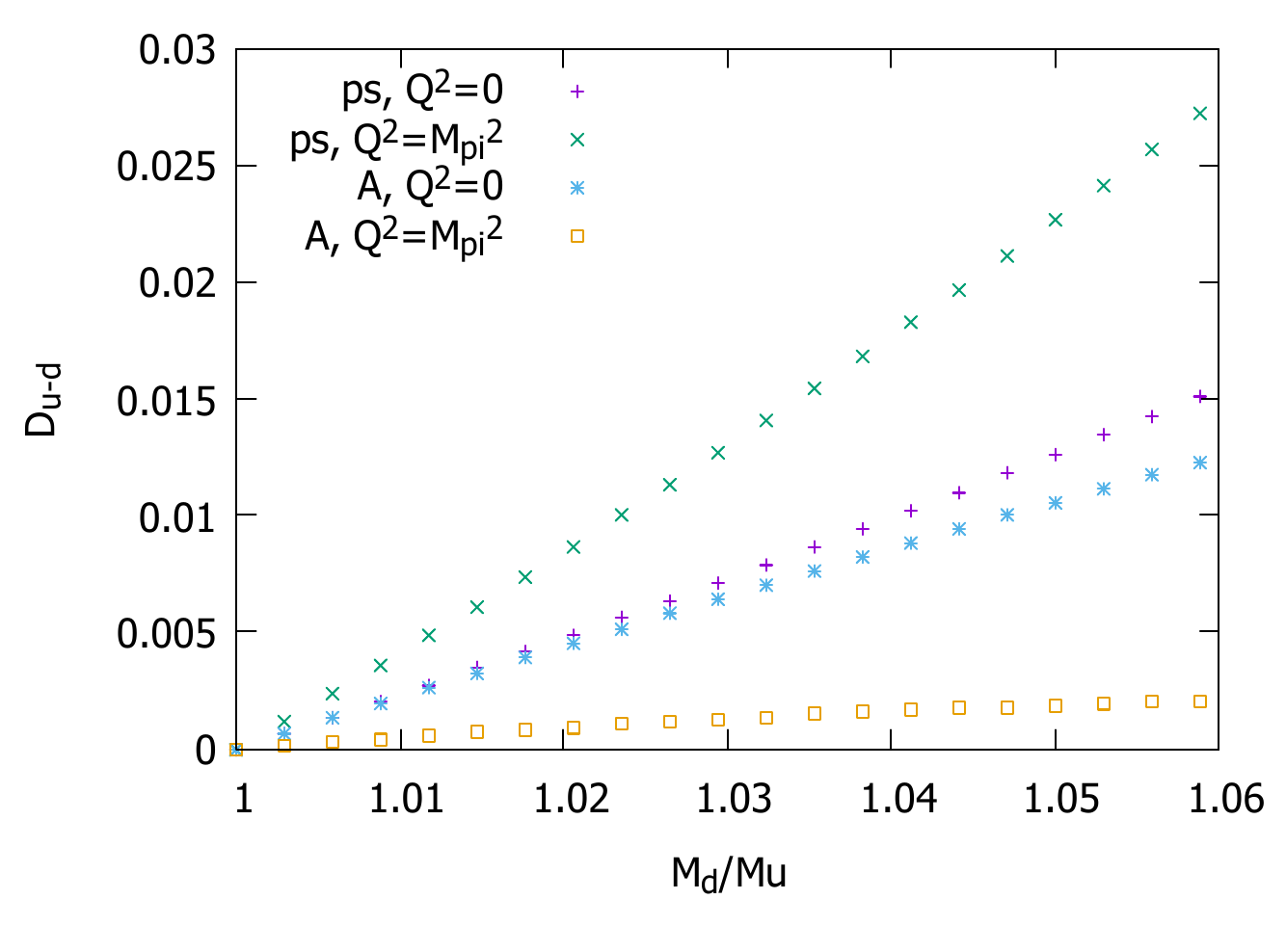}
\caption{ \label{fig:Diffud}
\small
Dependence of the difference of charged pion coupling constants
\eqref{Diffuddu} as a function of the mass ratio $M_d/M_u$, for 
the cases $Q^2=0$ and $Q^2=M_{\pi^\pm}^2$, with  $K^2=M_f^2$.
 }
\end{figure}

As noted before, the positive pion absorption
(negative pion emission) coupling constant in
both channels, $G_{du}^{\xi} (K^2 , Q^2) $ ($\xi=ps,A$),
might   be different from the 
negative pion absorption
(positive pion emission) coupling constant, 
$G_{ud}^{\xi}(K^2 , Q^2)$ in the case of nondegenerate 
quark masses.
The
difference between these two charged pion coupling constants, 
for both pseudoscalar and axial couplings,  is  shown
in Fig. \eqref{fig:Diffud}, by means of the following (on shell quarks)
ratios:
\begin{eqnarray} \label{Diffuddu}
D_{u-d} \equiv \frac{ G_{ud}^\xi (K^2=M_f^2,Q^2)  - G_{du}^\xi 
(K^2=M_f^2,Q^2) }{
 G_{ud}^\xi (K^2=M_f^2,Q^2)  + G_{du}^\xi  (K^2=M_f^2,Q^2) }.
\end{eqnarray}
Two cases are exhibitted,  $Q^2= M_{\pi^\pm}^2$
and $Q^2=0$.
It is seen that the variation is very small and 
basically   linear, i.e. goes 
 with the difference 
of masses $M_d-M_u$ in the leading order.
The relative variation of the axial coupling constant 
is smaller than the pseudoscalar.
The case of on shell or massive (off shell or massless) 
 pion  presents a larger variation 
for the pseudoscalar (axial) coupling constant.

\FloatBarrier

\subsection{
Pion coupling to Strange currents
}

\begin{figure}[ht!]
\centering
\includegraphics[width=100mm]{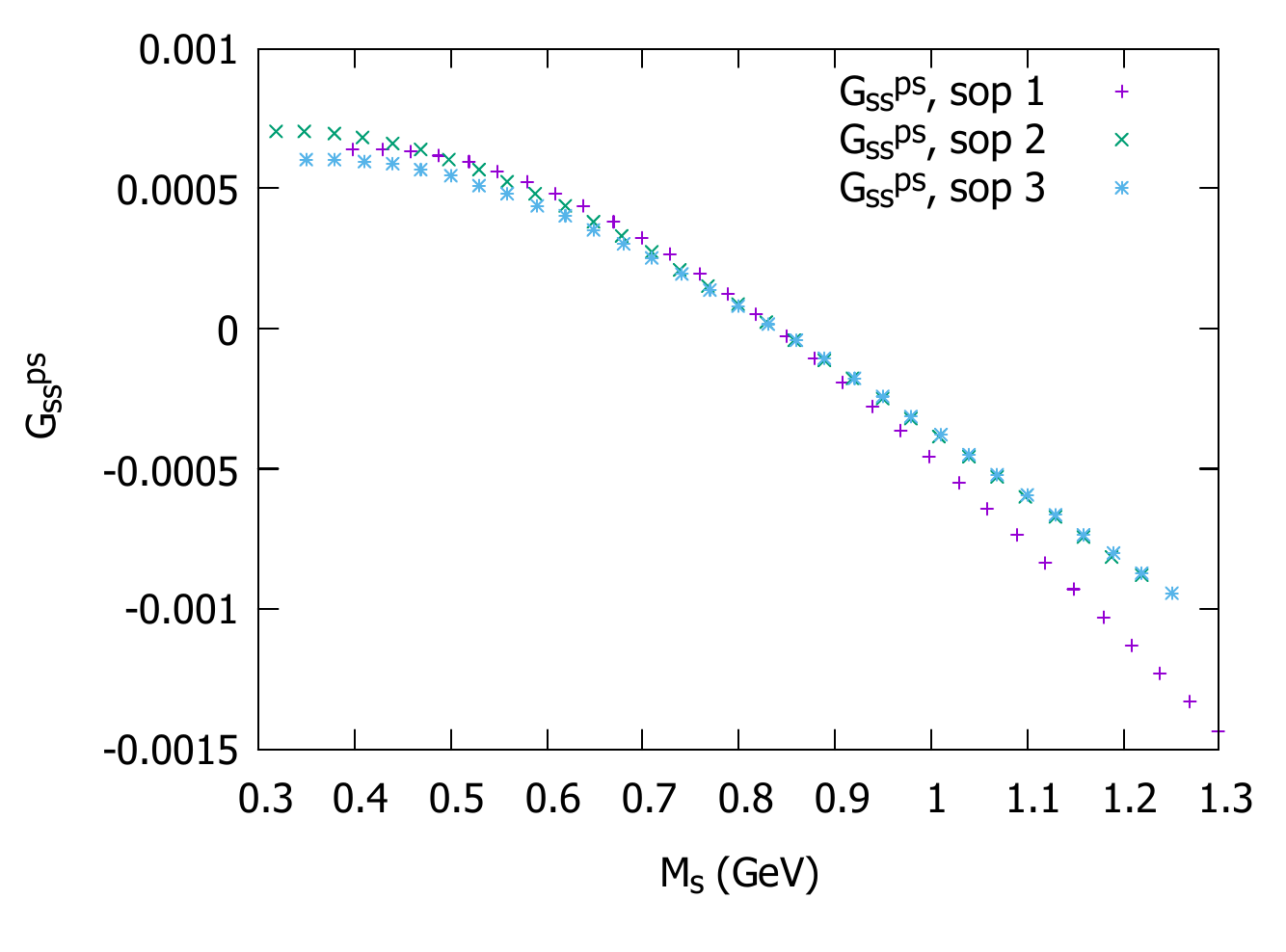}
\caption{ 
\label{fig:Gss-Gps}
\small
Ratio 
$G_{ss}^{ps}$ defined in 
\eqref{ratio-Gss} as a function of $M_s$ for the three s.o.p.
 }
\end{figure}

\begin{figure}[ht!]
\centering
\includegraphics[width=100mm]{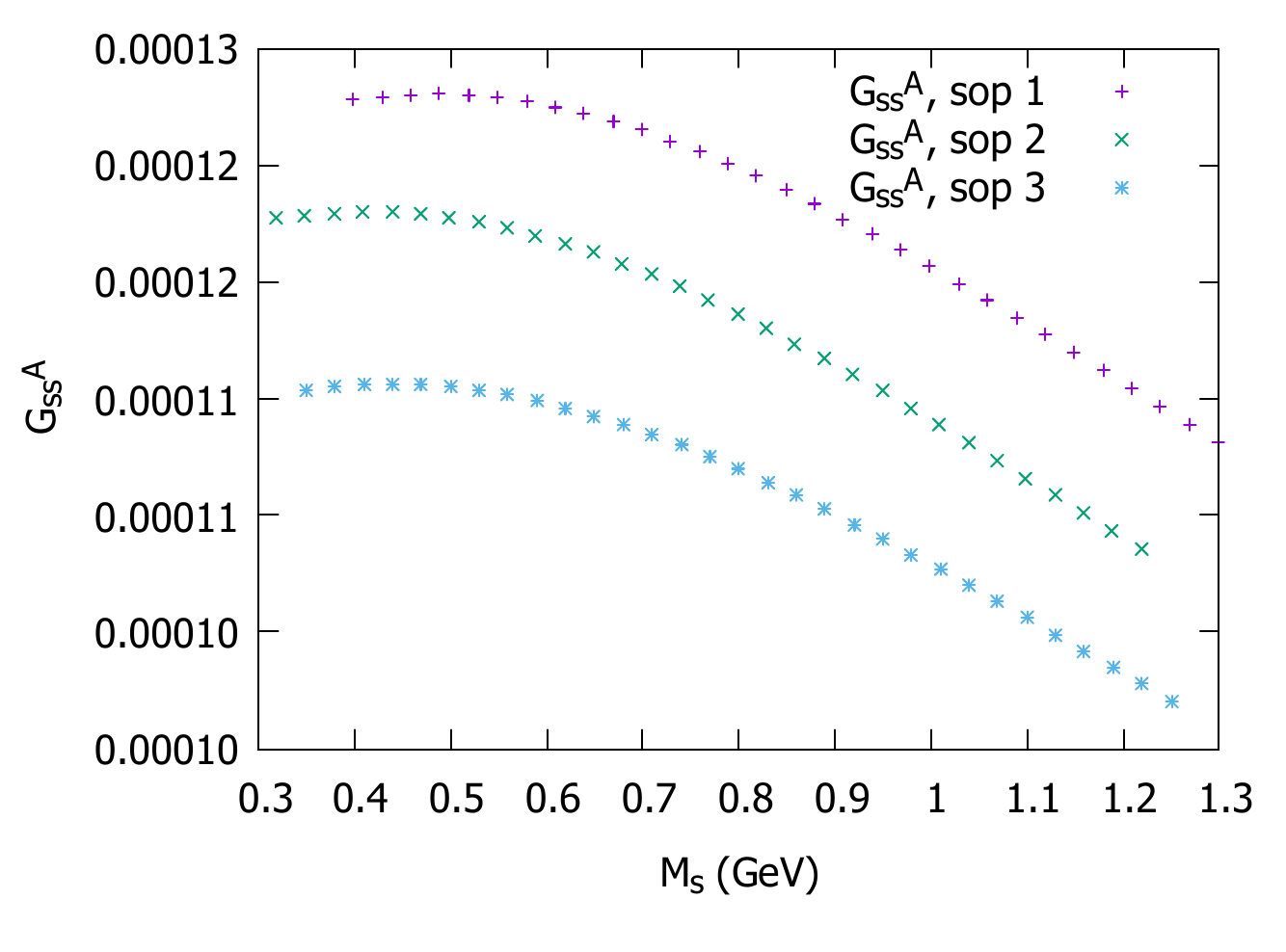}
\caption{ \label{fig:Gss-GA}
\small
Ratio 
$G_{ss}^{A}$ defined in 
\eqref{ratio-Gss} as a function of $M_s$ for the three s.o.p.
 }
\end{figure}

Both   pion couplings to strangeness current, the pseudo-scalar and the axial ones, are very small and they
 depend on the strange quark effective mass.
The behavior of the corresponding
coupling constants
will be shown with respect to the
pseudoscalar and to the axial coupling constants
(that are independent of
$M_s$)
by means of the following ratios with respect to the charged pion coupling constants with
definitions in Eq. \eqref{Gpi0mix}:
\begin{eqnarray}
\label{ratio-Gss}
G_{ss}^{ps} 
= \frac{G_{mix,ss}^{ps}}{ 
\frac{G_{ud}^{ps}+ G_{du}^{ps} }{2} },
\;\;\;\;\;\;\;
G_{ss}^{A} 
= \frac{G_{mix,ss}^{A}}{ 
\frac{G_{ud}^{A}+ G_{du}^{A} }{2} },
\end{eqnarray}
 where the coupling constants have been computed 
 for on shell quarks $K^2= M_f^2$
 and  off shell  pions
$Q^2 = 0$.
In Figs. \eqref{fig:Gss-Gps} and 
\eqref{fig:Gss-GA}
the ratios \eqref{ratio-Gss}
are exhibited as functions of the strange quark effective mass that is freely varied for each of the s.o.p.
 The pseudoscalar pion coupling to the strangeness current
 is relatively larger than 
 the axial pion coupling to 
 the corresponding strange quark current.
 Besides that,  $G_{mix,ss}^{ps}$ changes 
 to negative values for $M_s > 0.8$ GeV, for all the three  s.o.p.
 For $G_{ss}^{ps}$
 the s.o.p. with slightly different behavior is $A_1$
 that has larger effective masses
 whereas the
 relative behavior of $G_{ss}^{A}$
 is not really determined 
 by the strength of the
 effective masses of 
 each of the s.o.p.

\FloatBarrier

\FloatBarrier

\section{ Summary and conclusions}

Pseudoscalar and axial charged and neutral 
pion -constituent quark coupling constants 
were derived and  investigated   by considering 
quark mass non-degeneracy within a dynamical approach
 proposed previously by the author.
In this approach, coupling constants and form factors
are obtained from a quark determinant in the presence of 
meson fields and quark currents.
 Gluon effects are considered by 
means of a gluon effective propagator that dresses
background quark currents leading to  constituent quark 
currents.
These coupling constants
were calculated numerically 
for different sets of parameters of constituent quark
masses and in different kinematical points: for
  on shell or off shell pions and constituent quarks.
The pseudoscalar coupling constant calculated 
for off shell (or massless) charged pion and on shell
constituent quark  was fixed
as a renormalization condition by 
identifying it to a phenomenological value
$G^{ps}_{\pi^\pm} (K^2=M_f^2, Q^2=0) =13$.
With this condition,
 the resulting axial coupling constant is obtained with 
values close to the value adopted for the  
CQM, i.e. $g_A \sim 1$.
Being all the coupling constants functions of the quark constituent masses (and not the small 
current quark masses), 
the Goldberger Treiman relation - as usually defined -
 at the 
quark level may present,  in fact, a  large correction
for the values $G_{ps}=13$
and $g_A=1$.
Results suggest that lower values of quark effective masses
lead to smaller differences   between the 
neutral and charged pion couplings (both axial and pseudoscalar).
The differences between charged and neutral pion 
couplings to   nucleons 
recently obtained in different approaches \cite{granada,bochum,ArriolaAmaroPerez}
are  small  and this trend is reproduced with 
the constituent quark couplings in this work.
The specific hierarchies of the different coupling constants, however, are, in part, different.
Some of the differences between the 
values are so small that can be considered 
inside the error bars.
The  discrepancy appears basically when comparing 
the neutral pion coupling to protons or neutrons, 
or correspondingly
to up or down constituent quarks.
The  
average of neutral pion couplings to quark currents $\bar{u}u$  
and $\bar{d}d$ were compared to neutral pion coupling, respectively,
 to  
protons and neutrons.
 This comparison suggests
mixing interactions should be stronger  than those provided by flavor 
symmetry breaking   to reproduce values from 
Refs.\cite{granada,bochum}.
This extra mixing could be provided by 
instanton induced 
't Hooft-like interactions.
A more extensive investigation of the role of the 
specific 
gluon propagator will be provided elsehwere.

The resulting charged  
pion coupling constants  from
the quark determinant present full U(1) and CP invariances
with $G_{11}=G_{22}$.
It corresponds to 
an average of emitted and absorbed (positive or negative) pion,
being therefore the same for each charge eigenstate or 
for each type of process (emission or absorption).
For non-degenerate quark masses, there appears a small 
difference between
absorption and emission of a positive or negative pion.
This difference is proportional - in the leading order -
to the quark mass difference $(M_d- M_u)$.
However, 
high precision experimental results/fittings 
would be needed to identify such difference.

FSB induced interactions, that mix neutral flavor eigenstates,
 $i,j=0,3,8$,
leading to usual $\pi^0-\eta-\eta'$ mixing,
 have   been taken into 
account  for 
both the quark currents and for the neutral pion.
The shifts in the neutral pion 
 coupling to the  quark current $\bar{u}u$ 
are opposite to the  shifts in 
the coupling to the current $\bar{d}d$.
Besides that, a quite weak 
neutral pion coupling 
to the 
strange current $\bar{s}s$ was obtained for 
 both axial and pseudoscalar cases.
The relative strength of these couplings 
to strange current with 
respect to the pseudoscalar and axial ones, is 
very small  $\sim 10^{-3}, 10^{-4}$, 
so that their contributions for 
neutral pion couplings  to strange baryons would 
receive very tiny shifts only possibly identified with high 
precision measurements.
However, note that this coupling is proportional to the mixing interaction  $G_{i\neq j}$
that must receive contributions from 't Hooft-type interaction and sixth order 
interactions and these might increase considerably its value.
Nevertheless, given that hadrons are strongly interacting systems,
small contributions might, possibly, lead to sizeable effects.
Although the strength of the flavor dependent constituent quark
coupling constants might follow an hierarchy
   the behavior of these coupling constants by
inserting these couplings into the  
three-quark bound system for the nucleon  
can bring further insights into this subject.
Besides that, at finite temperature, baryon density or external 
electromagnetic fields, all these couplings must receive non-negligeable
contributions.
This type of investigation might seem not appropriated
for the understanding of hadron interactions because quarks are confined in
hadrons. 
However it is important to emphasize  that confinement 
operates in color degrees of freedom whereas flavor-isospin
 cannot be 
(completely)
confined, otherwise all the quark-based description of hadrons, and maybe QCD, 
would be invalidated.

 \vspace{0.5cm}

\centerline{\bf Acknowledgements}

The author thanks short discussion with C.D. Roberts in early 
stages of this work.
F.L.B. is a member of
INCT-FNA,  Proc. 464898/2014-5
and  he acknowledges partial support from 
 CNPq-312750/2021-8 and CNPq-407162/2023-2.

\numberwithin{equation}{section}

\appendix

%
%
%
%
%
%
%
%
%
%
%
%

\section{ Some relations}

\label{sec:appB}

%

%

The following flavor-structures compose 
the coupling constants/form factors $G^{\xi}_{ij}$ in \eqref{Gij-ff}
\begin{eqnarray} \label{G11}
G_{11}^\xi &=&    (  G_{u d}^\xi + G_{d u}^\xi  ),
\\
G_{33}^\xi &=&   ( G_{uu}^\xi  + G_{dd}^\xi  ),
\\
\label{Gmixings}
G_{03}^\xi  &=& \sqrt{\frac{2}{3}} (G_{uu}^\xi  - G_{dd}^\xi  ),
\\
\label{Gmixings2}
G_{08}^\xi  &=& \sqrt{\frac{2}{3} } ( G_{uu}^\xi + G_{dd}^\xi - 2  G_{ss}^\xi ),
\\
\label{Gmixings3}
G_{38}^\xi  &=& \frac{1}{\sqrt{3}} ( G_{uu}^\xi  - G_{dd}^\xi  ),
\\
G_{00}^\xi  &=&  \frac{2}{3} (G_{uu}^\xi + G_{dd}^\xi   + G_{ss}^\xi  ),
\\
\label{G88}
G_{88}^\xi  &=& \frac{1}{3} ( G_{uu}^\xi  +  G_{dd}^\xi  + 4 G_{ss}^\xi ).
\end{eqnarray}


\begin{thebibliography}{00}
 
%
%
%
%
%

\bibitem{CQM}
M. Lavelle, D. McMullan, 
Constituent quarks from QCD,
 Phys. Rept. {\bf 279}, 1  (1997).

\bibitem{CQM2} 
A. Manohar and H. Georgi, Nucl. Phys. B234, 189 (1984).
W. Plessas, The constituent-quark model nowadays, Int. J.
Mod. Phys. A 30, 1530013 (2015).


\bibitem{massgeneration1}
Y. B. Yang, J. Liang, Yu-J. Bi, Y. Chen, T. Draper, K. F. Liu,
and Z. Liu, Proton Mass Decomposition from the QCD
Energy Momentum Tensor, Phys. Rev. Lett. 121, 212001
(2018), and references therein.
C. Lorc\'e,  On the hadron mass decomposition. Eur. Phys. J. C 78, 120 (2018).



\bibitem{chengLi-IZ}
C. Itzykson, J.B. Zuber, Quantum Field Theory, 
McGraw-Hill, 1985.
T.P. Cheng, L.F. Li, 
Gauge theory of elementary particle physics,
Oxford, 1984. 


\bibitem{PCAC-gunnar}
G.S. Bali et al,
Solving the PCAC puzzle for nucleon axial and pseudoscalar form factors,
Phys. Lett. B 789, 666 (2019).

\bibitem{granada}
R. Navarro Pérez, J. E. Amaro, and E. Ruiz Arriola,
Precise determination of charge-dependent pion-nucleon-nucleon coupling constants,
Phys. Rev. C 95, 064001 (2017).




\bibitem{bochum}
P. Reinert , H. Krebs , and E. Epelbaum,
Precision Determination of Pion-Nucleon Coupling Constants Using
Effective Field Theory,
Phys. Rev. Lett. 126, 092501 (2021).

\bibitem{leutwyler}
J. Gasser  and H. Leutwyler,  Quark masses Phys. Rep. 87,
 77 (1982).

\bibitem{Donoghue}
J. F. Donoghue, Light quark masses and chiral symmetry,
Annu. Rev. Nucl. Part. Sci. 39, 1 (1989).


\bibitem{PDG}
K Nakamura
 et al (Particle Data Group) 2010 Review of particle physics J. Phys. G: Nucl. Part.
Phys. 37, 075021 (2010).
M. Tanabashi,
 et al (Particle Data Group) 2018 Phys. Rev. D 98 030001 (2018)

\bibitem{ISB-piN}
Martin Hoferichter, Bastian Kubis, Ulf-G. Meissner,
Isospin breaking in the pion–nucleon scattering lengths,
Phys. Lett. B678, 65 (2009).

\bibitem{weinberg-2010}
S. Weinberg, 
 Pions in Large 
N
 Quantum Chromodynamics,
Phys. Rev. Lett. {\bf 105}  (2010) 261601.   



\bibitem{EPJA-2016}
 F.L. Braghin,
Quark and pion effective couplings from
polarization effects, 
Eur. Phys. Journ.  A 52,  134  (2016).
F.L. Braghin,   Low energy constituent quark and pion effective couplings in
a weak external magnetic field, 
Eur. Phys.  J.   A 54, 45 (2018) .
ArXiv:1705.05926.

\bibitem{RMP2013}
C. A. Aidala,
S. D. Bass,
D. Hasch,
G. K. Mallot,
The spin structure of the nucleon,
Rev. Mod. Phys. 85, 655 (2013).


\bibitem{LeYaouanc1973}
A. Le Yaouanc, L. Oliver, O. Pene, and J.-C. Raynal,
"Naive" Quark-Pair-Creation Model of Strong-Interaction Vertices,
Phys. Rev. D8, 2223 (1973).


\bibitem{PRD-2019}
F.L. Braghin,
Pion constituent quark couplings strong form factors:
A dynamical approach,
Phys. Rev. D99, 01401 (2019).

\bibitem{PLB-2016}
F.L.Braghin, 
SU(2) Higher-order effective quark interactions from polarization,
Phys. Lett. B 761, 424 (2016).

\bibitem{ERV}
D. Ebert, H. Reinhardt, M.K. Volkov,
 Effective hadron theory of QCD, 
 Progr. Part. Nucl. Phys.
 33, 1  (1994).


\bibitem{GCM}
C. D. Roberts, R. T. Cahill, and J. Praschifka, The effective
action for the Goldstone modes in a global colour symmetry
model of QCD, Ann. Phys. (N.Y.) 188, 20 (1988).

\bibitem{meissner}
U.-G. Meissner, Low-energy hadron physics from effective
chiral Lagrangians, Phys. Rep. 161, 213 (1988).
B. Holdom, Approaching low-energy QCD with a gauged,
nonlocal, constituent-quark model, Phys. Rev. D 45, 2534
(1992).

\bibitem{chinese}
Ke Ren, Hui-Feng Fu, Qing Wang,
Derivation of the effective chiral Lagrangian for pseudoscalar, scalar,
vector, and axial-vector mesons from QCD,
Phys. Rev. D95, 074012 (2017).
 Q.Wang, Y.-P. Kuang, X.-L.Wang, and M. Xiao, Phys. Rev.
D 61, 054011 (2000)


 



\bibitem{feldmann-review}
T. Feldmann,  Quark structure of pseudoscalar mesons,
IJMPA, 
arXiv:hep-ph/9907491v2


\bibitem{instantons-ind}
M. Creutz, The ’t Hooft vertex revisited. Ann. Phys. (Amst.) 323,
2349 (2008)

\bibitem{PRD-2021}
F.L. Braghin,
Flavor-dependent U(3) Nambu–Jona-Lasinio coupling constant,
Phys. Rev. D103, 094028 (2021).


\bibitem{JPG-2022}
F.L. Braghin,
Strangeness content of the pion in the U(3)
Nambu–Jona–Lasinio model,
J. Phys. G 49, 055101 (2022).





\bibitem{EHSeff}
I.M.  Froldi, F.L. Braghin,
Large quark mass Euler-Heisenberg type action for QCD with quark sources,
arXiv:2406.19918.
 
\bibitem{PRD-2022a}
T.H. Moreira, F.L. Braghin,
Magnetic field induced corrections to the NJL model coupling constant
from vacuum polarization
Phys. Rev. D 105, 114009 (2022).

\bibitem{EPJA-2023}
W.F. de Sousa, F.L. Braghin,
U(5) Nambu–Jona–Lasinio model with flavor dependent coupling
constants: pseudoscalar and scalar mesons masses,
Eur. Phys. J. A 59, 271 (2023).



\bibitem{arriola}
E. Ruiz Arriola, Pion structure at high and low energies in chiral quark models, Acta Phys. Polon. B33, 4443 (2002), hep-ph/0210007.

\bibitem{HORN-ROB}
Craig D. Roberts,  Three Lectures on Hadron Physics,
J. Phys.: Conf. Ser. 706 022003 (2016).
Tanja Horn and Craig D Roberts, J. Phys. G: Nucl. Part. Phys. 43 073001 2016,
and references therein.



\bibitem{mixing1}
T. Feldmann,  P. Kroll and B. Stech, Mixing and decay constants of pseudoscalar mesons
Phys. Rev. D 58 114006 (1998).
P. Kroll , Mixing of pseudoscalar mesons and isospin symmetry breaking Int. J. Mod. Phys. A
20 331 (2005).

\bibitem{mixing2}
G.S. Bali, V. Braun, S. Collins , A. Schafer and J. Simeth, Masses and decay constants of the $\eta$ and $\eta'$ mesons
from lattice QCD, JHEP08, 137 (2021); (arXiv:2106.05398v1 [hep-lat]). 
F. J. Gilman, R. Kaufman, $\eta-\eta'$ mixing
angle, Phys. Rev. D36, 2761 (1987); 37(E) 3348 (1988). F. Ambrosino, et al (KLOE Colaboration), JHEP 07,
105 (2009).

\bibitem{GPS}
M. R. Schindler, T. Fuchs, J. Gegelia,  S. Scherer,
Axial, induced pseudoscalar, and pion-nucleon form factors in
manifestly Lorentz-invariant chiral perturbation theory,
Phys. Rev. C75, 025202 (2007).


\bibitem{PRD-2018a}
F.L. Braghin,  Light vector and axial mesons effective couplings to constituent quarks,
Phys. Rev.  D 97, 054025 (2018).


\bibitem{rijken} 
Th. A. Rijken, 
Constituent Quark Model and Nucleon-Nucleon Potentials,
arXiv[nucl-th]:2412.19858v1.
Th.A. Rijken, Y. Yamamoto,
Quark-Quark and Quark-Nucleon Potential Model
Extended-soft-core Meson-exchange Interactions,
arXiv[nucl-th]:2412.15732v1.


\bibitem{cornwall} J. M. Cornwall,
Entropy, confinement, and chiral symmetry breaking,
Phys. Rev. {\bf D 83}, 076001 (2011).




\bibitem{SDE}
P. Maris, P.C. Tandy,
Bethe-Salpeter study of vector meson masses and decay constants,
 Phys. Rev. C 60, 055214 (1999).
D. Binosi, L. Chang, J. Papavassiliou, C.D. Roberts, Bridging a gap between continuum
-QCD and ab initiopredictions
of hadron observables, Phys. Lett. B 742, 183 (2015) and references therein.
 A. Bashir, et al., Collective Perspective on Advances in Dyson-Schwinger Equation QCD, Commun. Theor.
Phys. 58,79 (2012).






\bibitem{CQM-baryon}
C. Schuren, E. Ruiz Arriola, K. Goeke,
Explicit chiral symmetry breaking in the Nambu-Jona-Lasinio model,
Nucl. Phys. A547, 612 (1992).
Amir H. Rezaeian, Niels R. Walet, Michael C. Birse,
Baryon structure in a quark-confining non-local NJL model,
Phys. Rev. C70, 065203 (2004).
Craig D. Roberts, Three lectures on Hadron Physics, 
Journ. of Physics: conference series 706, 022003 (2016).




\bibitem{ArriolaAmaroPerez}
E. Ruiz Arriola, J.E. Amaro, R. Navarro Perez, 
Three pion nucleon coupling constants,
Mod. Phys. Lett. A31, 28 (2016); nucl-th/1606.02171.




\bibitem{weinberg}
S, Weinberg,   Strong interactions at low energies. In: Bernstein, A.M., Holstein, B.R. (eds) Chiral Dynamics: Theory and Experiment. Lecture Notes in Physics, vol 452, 
Springer, Berlin, Heidelberg (1995).







\end{thebibliography}
\end{document}